\documentclass [aip, pop, amsmath, amssymb, reprint] {revtex4-1}

\usepackage{graphicx}
\usepackage{amsmath}
\usepackage{color}
\usepackage{verbatim}
\usepackage{setspace}
\usepackage{subfigure}
\DeclareMathAlphabet{\mathpzc}{OT1}{pzc}{m}{it}

\newcommand{\pfrac}[2]{\frac{\partial #1}{\partial #2}}
\newcommand{\pfraca}[1]{\frac{\partial}{\partial #1}}
\newcommand{\mvec}[1]{\mathbf{#1}}

\newcommand {\apgt} {\ {\raise-.5ex\hbox{$\buildrel>\over\sim$}}\ }
\newcommand {\aplt} {\ {\raise-.5ex\hbox{$\buildrel<\over\sim$}}\ }
\setkeys{Gin}{width=0.9\linewidth,keepaspectratio}

\begin{document}

\preprint{AIP/123-QED}

\title{Role of electron inertia and electron/ion finite Larmor radius
  effects in low-beta, magneto-Rayleigh-Taylor instability}

\author{B. Srinivasan}
\affiliation{Kevin T. Crofton Department of Aerospace and Ocean
  Engineering, Virginia Tech, Blacksburg, VA 24060}
\author{A. Hakim}
  \affiliation{Princeton Plasma Physics Laboratory, Princeton, NJ 
    08544}
\date{\today}

\begin{abstract}
  The magneto-Rayleigh-Taylor (MRT) instability has been investigated
  in great detail in previous work using magnetohydrodynamic and
  kinetic models for low-beta plasmas.  The work presented here
  extends previous studies of this instability to regimes where
  finite-Larmor-Radius (FLR) effects may be important.  Comparisons of
  the MRT instability are made using a 5-moment and a 10-moment
  two-fluid model, the two fluids being ions and electrons.  The
  5-moment model includes Hall stabilization whereas the 10-moment
  model includes Hall and FLR stabilization.  Results are presented
  for these two models using different electron mass to understand the
  role of electron inertia in the late-time nonlinear evolution of the
  MRT instability.  For the 5-moment model, the late-time nonlinear
  MRT evolution does not significantly depend on the electron inertia.
  However, when FLR stabilization is important, the 10-moment results
  show that a lower ion-to-electron mass ratio (i.e. larger electron
  inertia) under-predicts the energy in high-wavenumber modes due to
  larger FLR stabilization.
\end{abstract}

\maketitle

\section{Introduction}

The Rayleigh-Taylor instability (RTI) is an instability of a fluid
interface appropriately oriented in the presence of an acceleration.
Conventional hydrodynamic \cite{dimonte_2000, dimonte_2004} and
magnetohydrodynamic (MHD) \cite{chandrasekhar_book_1961} simulations
have been performed to study the growth of this instability for
regimes ranging from space and astrophysical plasmas \cite{jun_1995,
  stone_2007} to fusion plasmas \cite{atzeni_2004, srinivasan_2013}.
The single-fluid MHD model does not alter the growth rate of
2-dimensional RTI when a purely out-of-plane magnetic field is present
except due to large thermal conduction and diffusion mechanisms.
Hence, a single-fluid ideal-MHD model only produces an altered RTI
growth compared to a neutral fluid model if there is an in-plane
magnetic field that can stretch, twist, and fold due to the MHD dynamo
\cite{childress_book_1995, srinivasan_2013}.  The Hall-MHD model,
however, can significantly alter the growth of the RTI even in the
presence of a purely out-of-plane magnetic field \cite{huba_1998}.
The evolution of RTI in the presence of non-ideal effects has been
less studied with significant early advances made by Huba et al
\cite{huba_1987, huba_1996, huba_1998} to illustrate the role of Hall
terms and finite Larmor radius (FLR) effects.  Recent work has studied
the role of non-ideal effects on RTI growth for inertial confinement
fusion (ICF) regimes in the presence of magnetic fields using Hall-MHD
models with results compared to experiments \cite{manuel_2012,
  srinivasan_2012, gao_2012, srinivasan_2013, srinivasan_epl_2014}.
First observations of magneto-Rayleigh-Taylor (MRT) instability
evolution in the presence of magnetic and viscous effects have been
made in recent experiments \cite{adams_2015}.  Other recent work has
explored the development of secondary instabilities in single-mode RTI
growth using a fully kinetic code \cite{umeda_2016} and the role of
non-MHD effects in MHD scale RTI growth \cite{goto_2015, umeda_2017}.
These works have explained the importance of Hall and gyro-viscous
effects on RTI growth and how the presence of these additional terms
introduces secondary instabilities that alter the overall evolution of
the RTI.

The presence of the Hall term can increase the growth rate of RTI
significantly even with a purely out-of-plane magnetic field
particularly for high-wavenumber modes.  On the other hand, FLR
effects can have a significant stabilizing effect on RTI growth
\cite{huba_1996} with the potential to completely stabilize
high-wavenumber modes.  Here, the role of electron inertia and FLR
effects is studied using five moment and ten moment two-fluid
models\cite{wang_2015} with comparisons of mode growth to the Hall-MHD
and hybrid simulations of Ref.[\onlinecite{huba_1998}].  As far as the
authors are aware, ten-moment two-fluid simulations of RTI have not
been studied previously.  Ref.[\onlinecite{huba_1998}] describes the
regimes of RT growth when each of Hall physics and FLR physics is
included and the associated stabilization.  In this work, simulations
are performed in 2-dimensions using purely out-of-plane magnetic
fields in the low-beta regime with the five moment and ten moment
models.  Any in-plane magnetic field components provide stabilization
of short-wavelength modes due to the MHD dynamo even in
ideal-MHD. Hence, they are neglected in this work. The five moment
model includes electron inertia effects with Hall physics but no FLR
effects.  The ten moment model includes electron inertia effects, Hall
physics, and FLR effects, with a closure for the gradient of the
heat-flux tensor. This work presents results showing that electron
inertia alters the RTI growth rate in regimes where the FLR effects
are important. This can have significant consequences for problems
where an artificial ion-to-electron mass ratio is used.  It is shown
that the FLR stabilization of the RTI is dominated by the ion FLR
effects and the ion-to-electron mass ratio alters RTI growth in
regimes where FLR effects are important.

\section{Problem setup} \label{sec:setup}

The setup of interest in this work is a 2-dimensional planar domain
with a density gradient on a magnetized plasma in the presence of
gravity such that it is RT unstable when perturbed.  An out-of-plane
magnetic field is included.  The inclusion of an out-of-plane magnetic
field does not alter the growth of the planar RTI from the
unmagnetized case when a single-fluid MHD model is used.  However, the
inclusion of Hall effects and FLR effects can alter the stability
criteria for RTI in such a geometry \cite{huba_1998}.  What is not yet
well understood is the role of electron inertia and the full pressure
tensor evolution in the stability and dynamics of the magnetized RTI
particularly in the nonlinear phase of instability growth.  This work
uses a planar geometry in 2D to study these non-MHD effects in RTI
growth and evolution using five- and ten-moment two-fluid models that
include a complete description of the electron continuity, momentum,
and energy while also including the full pressure tensor description
in the ten-moment equations. Simulations are performed initializing a
low beta plasma with a purely out-of-plane magnetic field.

The parameter space spans a regime where FLR effects are significant
similar to the description in Ref.[\onlinecite{huba_1998}].  The
initial conditions use an Atwood number of $0.17$, an acceleration of
$0.09$, an initial plasma beta of $0.071$, an ion-to-electron
temperature ratio of $0.1$ with a hyperbolic tangent density profile
and an initial pressure balance.  The domain size is $3\delta_i$ in
the x-direction and $3.75\delta_i$ in the y-direction, with $\delta_i$
representing the ion skin depth.  The perturbation is initially
applied to the ion and electron densities to initialize modes $3$ to
$32$ \cite{srinivasan_followup_2012}.  The grid resolution used for
all simulations is $1500\times 1875$ so as to sufficiently resolve all
initially perturbed modes consistently across the different models.

Previous single-mode RTI simulations have been performed in a similar
planar geometry using an artificial ion-to-electron mass ratio of
$25$, which those authors chose for computational tractability
\cite{umeda_2016, umeda_2017}. However, as shown here, an artificially
high electron mass can significantly alter the growth and dynamics of
RTI particularly in the late nonlinear phase of multimode RTI. For the
parameter regimes chosen, FLR effects are important and FLR
stabilization is expected for selected modes. Simulations are
performed to isolate ion and electron FLR effects and their individual
roles in the growth and dynamics of the magneto-RTI.

\section{Equations and numerical method}

A two-fluid plasma model is used in this work and is derived by taking
moments of the Boltzmann equation and treating the electrons and ions
as two separate fluids.  The five-moment equation system
\cite{shumlak_2003} results by assuming an isotropic pressure with an
ideal gas equation of state.  This produces Euler equations for each
of the electron and ion fluids with resulting equations having
homogeneous, hyperbolic parts and inhomogeneous source
terms. Retaining the tensor of the second moment provides the
ten-moment equations \cite{hakim_2008} which includes the full
anisotropic pressure tensor with a local collisionless heat flux
closure.  Maxwell's equations are used to evolve the electromagnetic
terms.  A finite volume scheme is used to solve the five- and
ten-moment equations \cite{hakim_2006} in the {\tt Gkeyll} code
\cite{ng_2015, cagas_2017, juno_2018}.  To allow realistic mass ratios
in the models without having the electron plasma and cyclotron
frequencies restrict the time-step, the hyperbolic part of the
equations are evolved explicitly using a single-step finite-volume
scheme while the source terms are treated with implicitly. Operator
splitting is used to solve the full system. These equation systems are
described below.

\subsection{Five-moment two-fluid plasma model}

The equations described here are the five-moment two-fluid (5M)
equations that result from taking the zeroth, first, and second
moments of the Vlasov equation and they are closed with an ideal gas
equation of state.  Isotropic pressure is assumed and collisions are
neglected. The electrons and ions are each described by the Euler
equations with source terms coupling the fluids and the fields. The 5M
equations are given by
\begin{align}
  \pfrac{\rho_s}{t} + \nabla\cdot(\rho_s\mvec{u}_s) &= 0 \label{eq:tfcont}\\
  \pfrac{\rho_s \mvec{u}_s}{t} +
  \nabla\cdot(\rho_s\mvec{u}_s\mvec{u}_s + p_s \mvec{I}) &=
  \frac{\rho_s q_s}{m_s}(\mvec{E}+\mvec{u}_s\times\mvec{B}) \label{eq:tfmom}\\
  \pfrac{\epsilon_s}{t} + \nabla\cdot((\epsilon_s+p_s)\mvec{u}_s) &=
  \frac{\rho_s q_s}{m_s}\mvec{u}_s\cdot\mvec{E} \label{eq:tfener}
\end{align}
where subscript, $s$, denotes electron or ion species. $q$ is the
charge, $m$ is the mass, $\rho$ is the mass density, $\mvec{u}$ is the
velocity, $\mvec{E}$ is the electric field, $\mvec{B}$ is the magnetic
field, $p$ is the pressure, and $\epsilon$ is the total energy. The
energy is defined as
\begin{align}
  \epsilon_s\equiv\frac{p_s}{\gamma -1}+\frac{1}{2}\rho_su_s^2.
\end{align}
Maxwell's equations are used to evolve the electric and magnetic
fields.
\begin{align}
  \pfrac{\mvec{B}}{t}+\nabla\times \mvec{E} &= 0 \label{eq:tfecurl} \\
  \frac{1}{c^2}\pfrac{\mvec{E}}{t}-\nabla\times \mvec{B} &=
  -\mu_0\mvec{J} \label{eq:tfbcurl} \\
  \nabla\cdot\mvec{E} &= \frac{\varrho_c}{\varepsilon_0} \label{eq:tfediv} \\
  \nabla\cdot\mvec{B} &= 0 \label{eq:tfbdiv}
\end{align}
where $\varrho_c$ and $\mvec{J}$ are the charge density and the
current density defined by
\begin{align}
  \varrho_c &\equiv \sum_s \frac{q_s}{m_s}\rho_s \label{eq:mecharge} \\
  \mvec{J} &\equiv \sum_s \frac{q_s}{m_s}\rho_s\mvec{u}_s. \label{eq:mecurrent}
\end{align}
To satisfy the divergence constraints of Eqs. \ref{eq:tfediv} and
\ref{eq:tfbdiv}, a hyperbolic form of Maxwell's equations is evolved
to include divergence corrections\cite{munz_2000}.  The fluids and
fields are coupled via source terms.  For a plasma of $S$ species, the
5M equations consist of $5S+8$ equations.

\subsection{Ten-moment two-fluid plasma model}

The ten-moment two-fluid (10M) equations are obtained by retaining the
anisotropic pressure tensor of the second moment of the Vlasov
equation. The following higher order moments are defined,
\begin{align}
  \mathcal{P}_{ij} &\equiv m \int v_i v_j f d\mvec{v}, \label{eq:mcp} \\
  \mathcal{Q}_{ijk} &\equiv m \int v_i v_j v_k f d\mvec{v}.
  \label{eq:mcq}
\end{align}
which are used to obtain the set of \emph{exact} moment equations
\cite{wang_2015}
\begin{align}
  &\pfrac{n}{t}+\pfraca{x_j}(nu_j) = 0, \label{eq:num} \\
  &m\pfraca{t}(nu_i) + \pfrac{\mathcal{P}_{ij}}{x_j}
  =
  nq(E_i+\epsilon_{ijk}u_jB_k), \label{eq:mom} \\
  &\pfrac{\mathcal{P}_{ij}}{t} + \pfrac{\mathcal{Q}_{ijk}}{x_k}
  =
  nqu_{[i}E_{j]}
  + \frac{q}{m}\epsilon_{[ikl}\mathcal{P}_{kj]}B_l. \label{eq:press}
\end{align}
The square brackets around indices represent the minimal sum over
permutations of free indices needed to yield completely symmetric
tensors. For example $u_{[i}E_{j]} = u_iE_j + u_jE_i$.
Equations\thinspace(\ref{eq:num})--(\ref{eq:press}) are 10 equations
(1+3+6) for 20 unknowns ($\mathcal{Q}_{ijk}$ has 10 independent
components). In general any finite set of exact moment equations will
always contain more unknowns than equations. Writing
\begin{align}
  \mathcal{Q}_{ijk} &= {Q}_{ijk} + u_{[i}\mathcal{P}_{jk]} - 2nm u_i u_j u_k \
  \label{eq:heat_ijk}
\end{align}
to close this system of equations we need a closure approximation for
the heat-flux, ${Q}_{ijk}$, defined as
\begin{align}
  {Q}_{ijk} &\equiv m \int (v_i-u_i) (v_j-u_j)(v_k-u_k) f d\mvec{v}.
\end{align}
Eqs.\thinspace(\ref{eq:num})--(\ref{eq:press}), closed with an
approximation for the divergence of the heat-flux tensor (along with
Maxwell equations), are the \emph{ten-moment two-fluid
  equations}\cite{hakim_2008}. For a plasma of $S$ species, they
consist of $10S+8$ equations.

The heat flux closure for the ten-moment two-fluid model allows
inclusion of the full pressure tensor unlike the five-moment two-fluid
model which assumes isotropic pressure.  Hence, the ten-moment
two-fluid model can capture agyrotropy of the pressure tensor.
Agyrotropy is a measure of the asymmetry of the pressure tensor about
a local magnetic field direction. This information is contained in the
full anisotropic pressure tensor. Several scalar measures of
agyrotropy have been developed. In the following, the measure
$\sqrt{Q}$, suggested by Swisdak\cite{swisdak_2016} is used
\begin{align}
  Q = 1 - \frac{4 l_2}{(l_1-P_\parallel)(l_1+P_\parallel)}
  \label{eq:agyro}
\end{align}
where $l_1 = P_{xx}+P_{yy}+P_{zz}$, $l_2 = P_{xx}P_{yy} + P_{xx}P_{zz}
+ P_{yy}P_{zz} - P_{xy}^2 - P_{xz}^2-P_{yz}^2 $, and $P_\parallel =
\hat{\mvec{b}}\cdot\mvec{P}\cdot\hat{\mvec{b}}$.  Comparisons have
been made between the agyrotropy calculated using a ten-moment model
versus a fully kinetic model for planar problems \cite{ng2017} and for
global magnetosphere problems \cite{Wang:2018dq}.  It is found that
ten-moment models produce lower values of agyrotropy compared to
kinetic results but the values are close in magnitude and the
structures are very similar. The motivation for using a ten-moment
model as opposed to a kinetic model in this work is the computational
effort associated with using realistic ion-to-electron mass ratios as
compared to the large computational cost of performing a 2X3V (2
spatial dimensions and 3 velocity dimensions) kinetic simulation with
a realistic mass ratio.  This work shows how agyrotropy changes with
different ion-to-electron mass ratios and consequently, how
ion-to-electron mass ratio alters RTI growth in regimes where FLR
effects are important.

\section{Results and Analysis}

\subsection{Five-moment versus ten-moment linear theory for RT growth} \label{sec:5M10M}

Simulations are performed using the five-moment and ten-moment models
for the initial conditions described in Sec.~\ref{sec:setup}.  The
chosen parameter space is in a regime where the ion FLR effects
dominate resulting in FLR stabilization of short-wavelength RTI.  The
five-moment model captures Hall effects but does not have sufficient
physics in the equations to capture FLR effects.  When using the
ten-moment model, which incorporates Hall effects along with FLR
effects, the FLR stabilization becomes dominant.

The Hall term influences RT growth based on \cite{huba_1998}
\begin{align}
  \frac{\omega^2}{k_yg} =
  \frac{-A+\omega/\Omega_i}{1-A\omega/\Omega_i}
\end{align}
where $A$ is the Atwood number, $g$ is the acceleration, and
$\Omega_i$ is the ion cyclotron frequency.  Including FLR effects
results in higher growth rates of RTI for lower-wavenumber modes as
compared to the 5M model and the RT dispersion is given by
\cite{huba_1998}
\begin{align}
  \omega^2 - k_y^2A\frac{cT}{eB} \omega + k_yg = 0
\end{align}
where $T$ is the temperature, $B$ is the magnetic field, $e$ is the
elementary charge, and $c$ is the light speed.  The FLR effects in the
regime described here completely stabilizes modes higher than
approximately 22 in the linear regime.

While the linear growth of Hall and FLR physics and the associated
stabilization has been studied previously, the effect of these terms
on nonlinear instability evolution remains to be understood.  Of
particular interest is the wavenumber spectrum that persists in the
nonlinear phase of the RT growth and the dependence of the nonlinear
behavior on the ion-to-electron mass ratio.

\subsection{Role of electron inertia in the five-moment model} \label{sec:5Mmassratio}

\begin{figure*}[!htb]
  \includegraphics[width=1.0\linewidth]{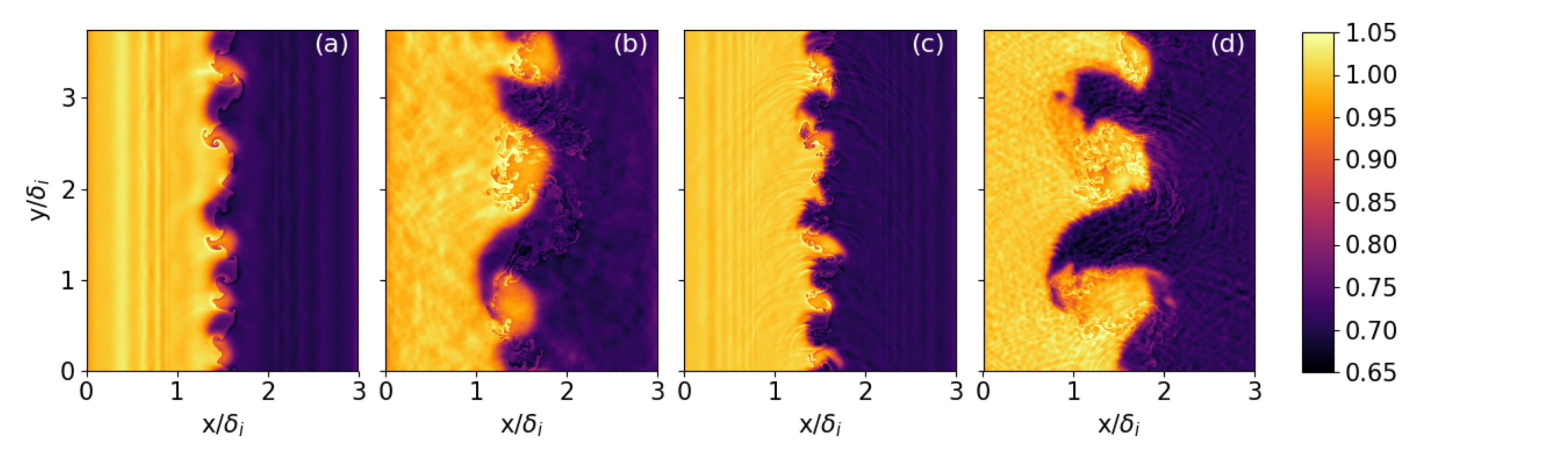}
  \caption{Ion densities after 10/$\omega_{ci0}$ (a) and
    35/$\omega_{ci0}$ (b) for ion-to-electron mass ratios $M=25$ and
    the corresponding densities at 10/$\omega_{ci0}$ (c) and
    35/$\omega_{ci0}$ (d) for $M=1836$ using the 5M model.  Note the
    similarity in the nonlinear RT growth of the artificial mass ratio
    using higher electron inertia compared to the realistic mass ratio
    case.}
  \label{fig:5Mdens}
\end{figure*}
\begin{figure*}[!htb]
  \includegraphics[width=0.45\linewidth]{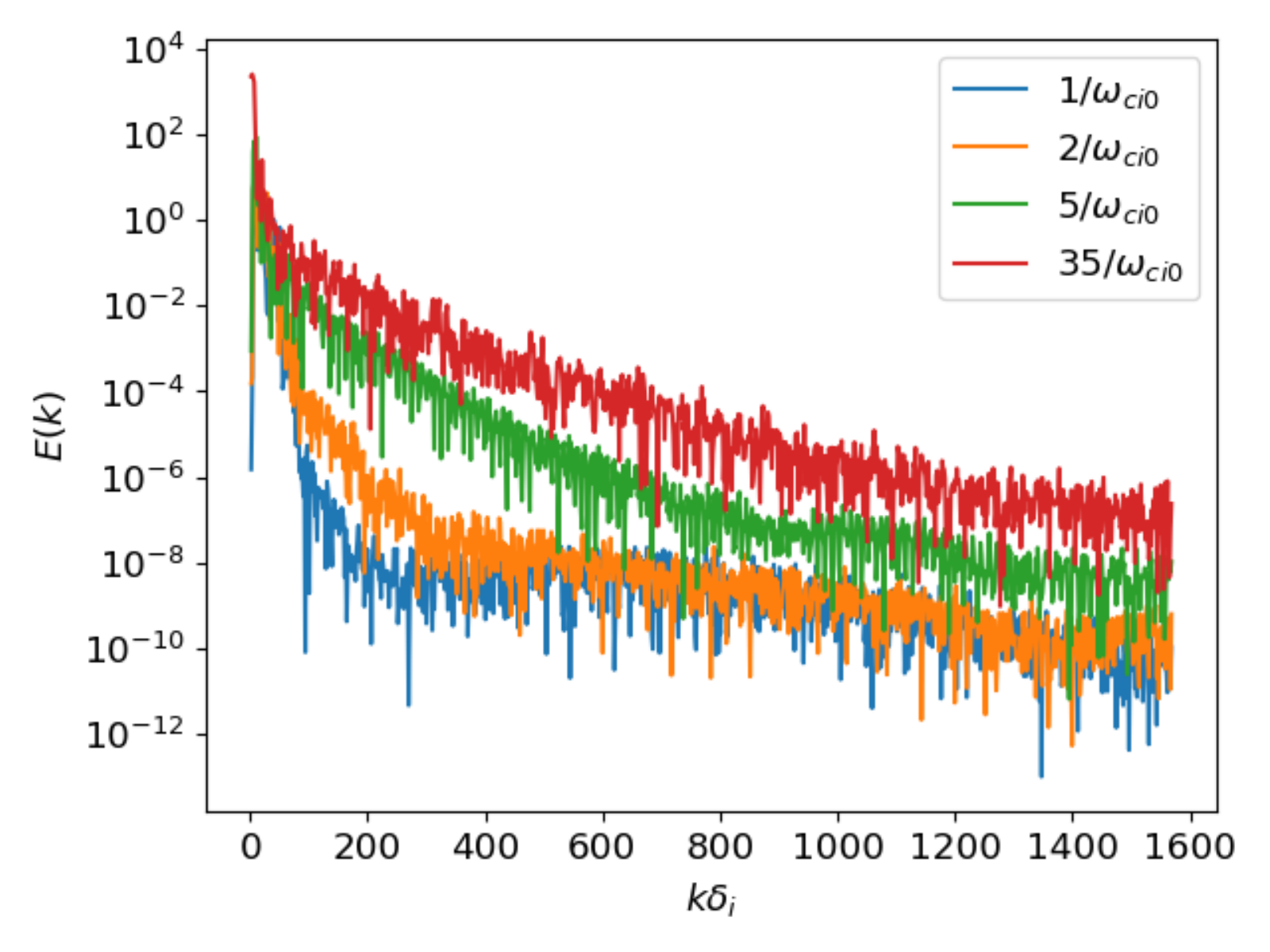}
  \includegraphics[width=0.45\linewidth]{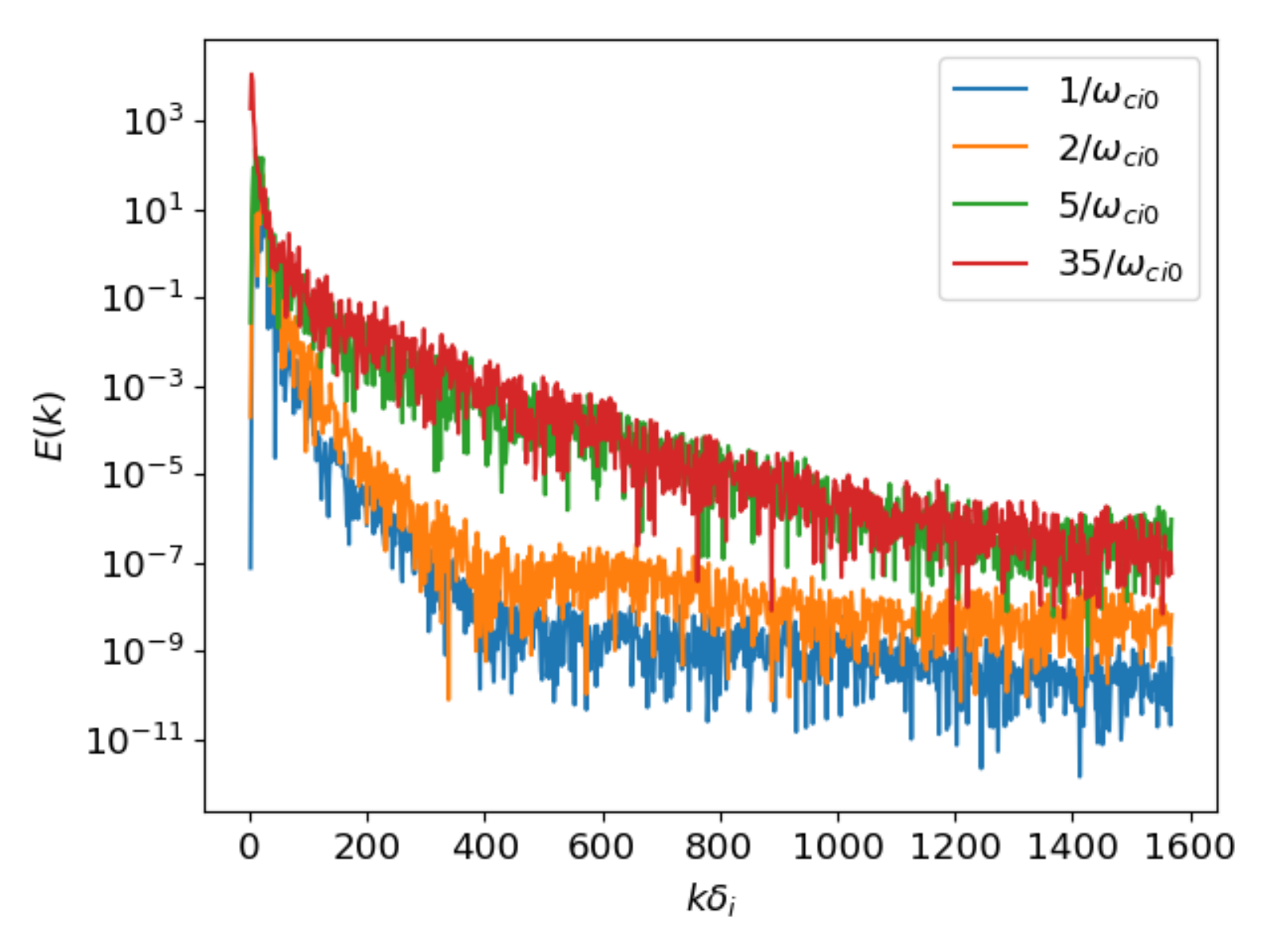}
  \caption{Power spectral density as a function of wavenumber for
    ion-to-electron mass ratios $M=25$ (left) and $M=1836$ (right)
    using the using the 5M model.  Solutions are shown for several
    different times during the evolution.  The ion density is
    integrated in the $x$-direction to obtain the spectrum in the
    $y$-direction.  Note that the late-time spectra of the two mass
    ratios are similar. }
  \label{fig:5Mfft}
\end{figure*}

Figure~\ref{fig:5Mdens} presents ion density at two separate times
during the evolution for multi-mode simulations with the 5M model
where modes $3$ to $32$ are initialized for two ion-to-electron mass
ratios of $25$ and $1836$.  The perturbation is of form described in
Refs.[\onlinecite{srinivasan_2012, srinivasan_followup_2012}].  The
subplots (a) and (b) of Fig.~\ref{fig:5Mdens} present the early-time
and late-time evolution of the ion density for a mass ratio of $25$
and subplots (c) and (d) present early- and late-time evolution for a
mass ratio of $1836$.  Both ion-to-electron mass ratio simulations
have higher-wavenumber modes growing early-in-time as seen in
Fig. 1(a) and 1(c) (after $10/\omega_{ci0}$) compared to late-in-time
in Fig. 1(b) and 1(d) (after $35/\omega_{ci0}$).  The higher
ion-to-electron mass ratio of $1836$ has slightly higher-wavenumber
modes and larger amplitudes compared to a mass ratio of $25$, but the
dominant growing modes are similar in both cases.  This is further
highlighted in the wavenumber spectrum for several different times
shown in Fig.~\ref{fig:5Mfft} for each of the two mass ratios
presented.  Note that the late-time power spectral densities (at
$35/\omega_{ci0}$) do not substantially differ between the two mass
ratios.  While the dominant modes that remain in the solution in the
nonlinear phase of instability growth are long-wavelength modes from
the qualitative results of Fig.~\ref{fig:5Mdens}, the energy across
all modes is approximately similar for the two mass ratios.  In this
regime with magnetized ions (low plasma beta), any stabilization of
higher wavenumber modes is due to the Hall term and the dominant
growing modes are of lower wavenumber \cite{huba_1998} as the 5M model
does not include FLR effects.

\subsection{Role of electron inertia in the ten-moment model}

\begin{figure*}[!htb]
  \includegraphics[width=1.0\linewidth]{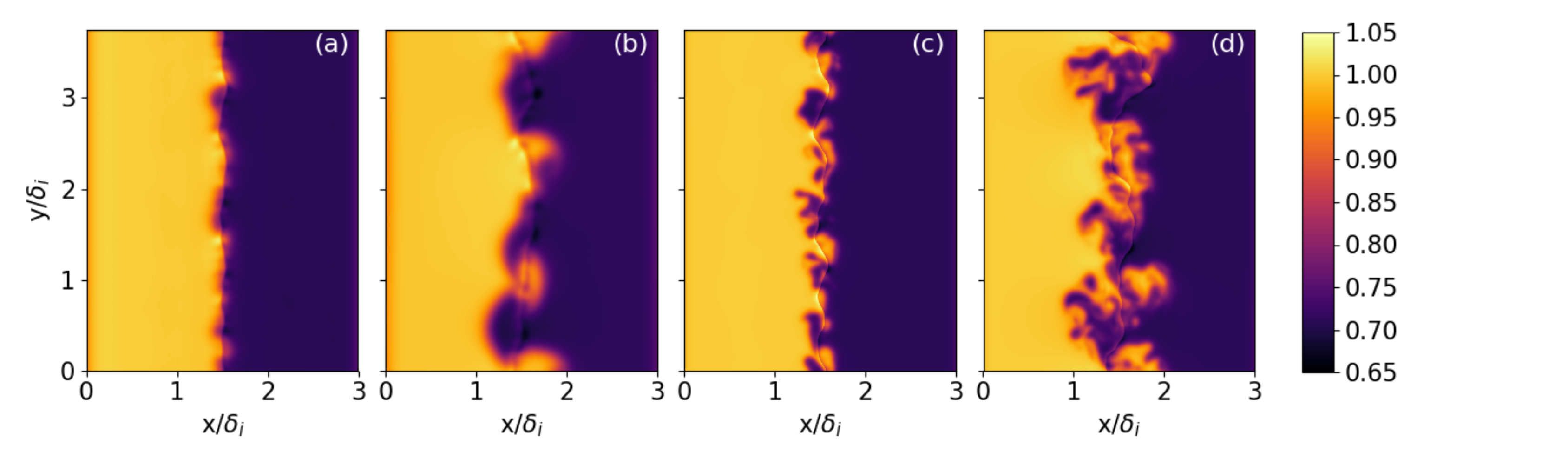}
  \caption{Ion densities after 10/$\omega_{ci0}$ (a) and
    35/$\omega_{ci0}$ (b) for ion-to-electron mass ratios $M=25$ and
    the corresponding densities at 10/$\omega_{ci0}$ (c) and
    35/$\omega_{ci0}$ (d) for $M=1836$ using the 10M model.  Note the
    higher wavenumber modes present in the nonlinear RT growth of the
    realistic (higher) mass ratio case compared with the artificial
    (lower) mass ratio case.}
  \label{fig:10Mdens}
\end{figure*}
\begin{figure*}[!htb]
  \includegraphics[width=0.45\linewidth]{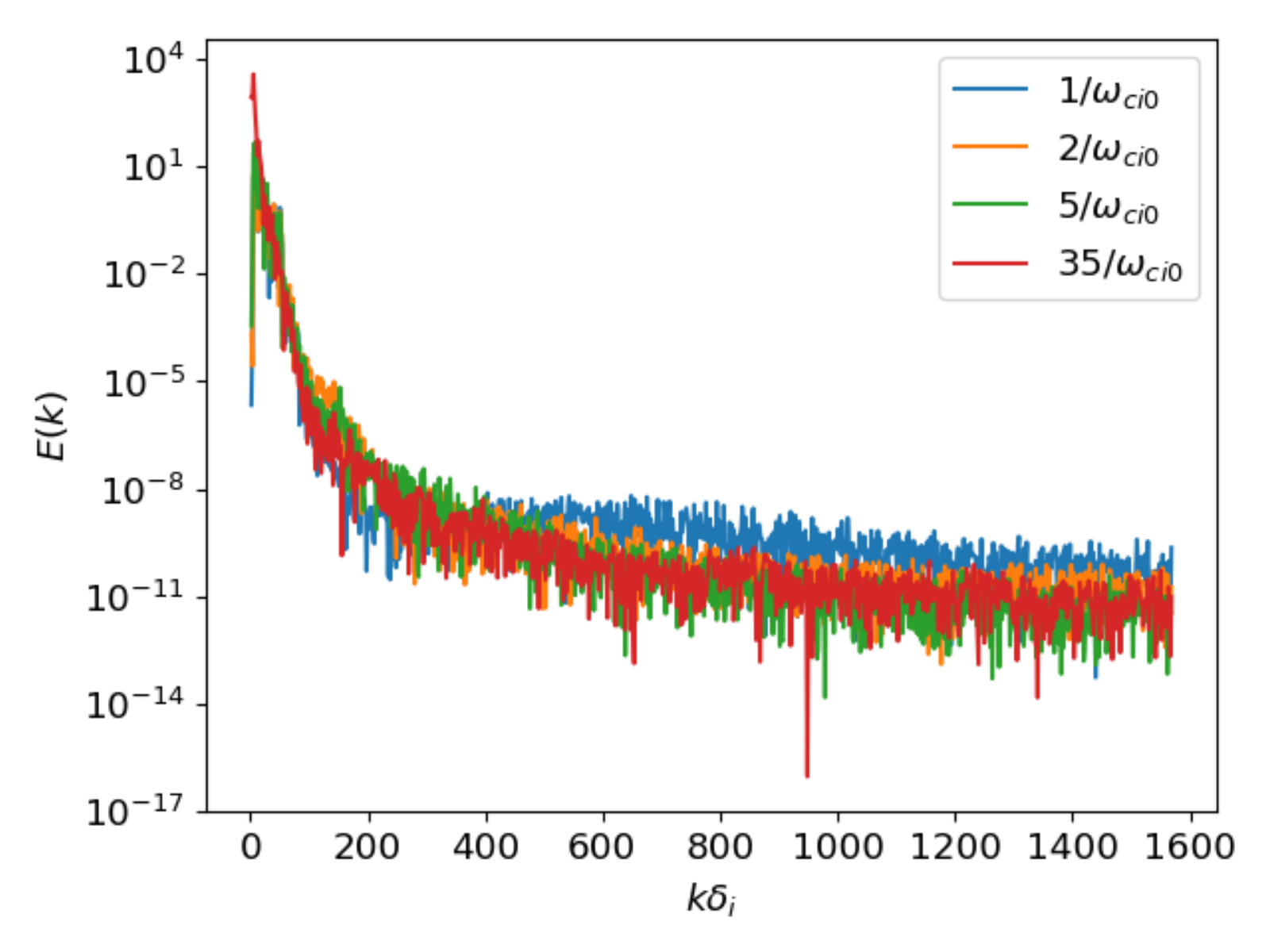}
  \includegraphics[width=0.45\linewidth]{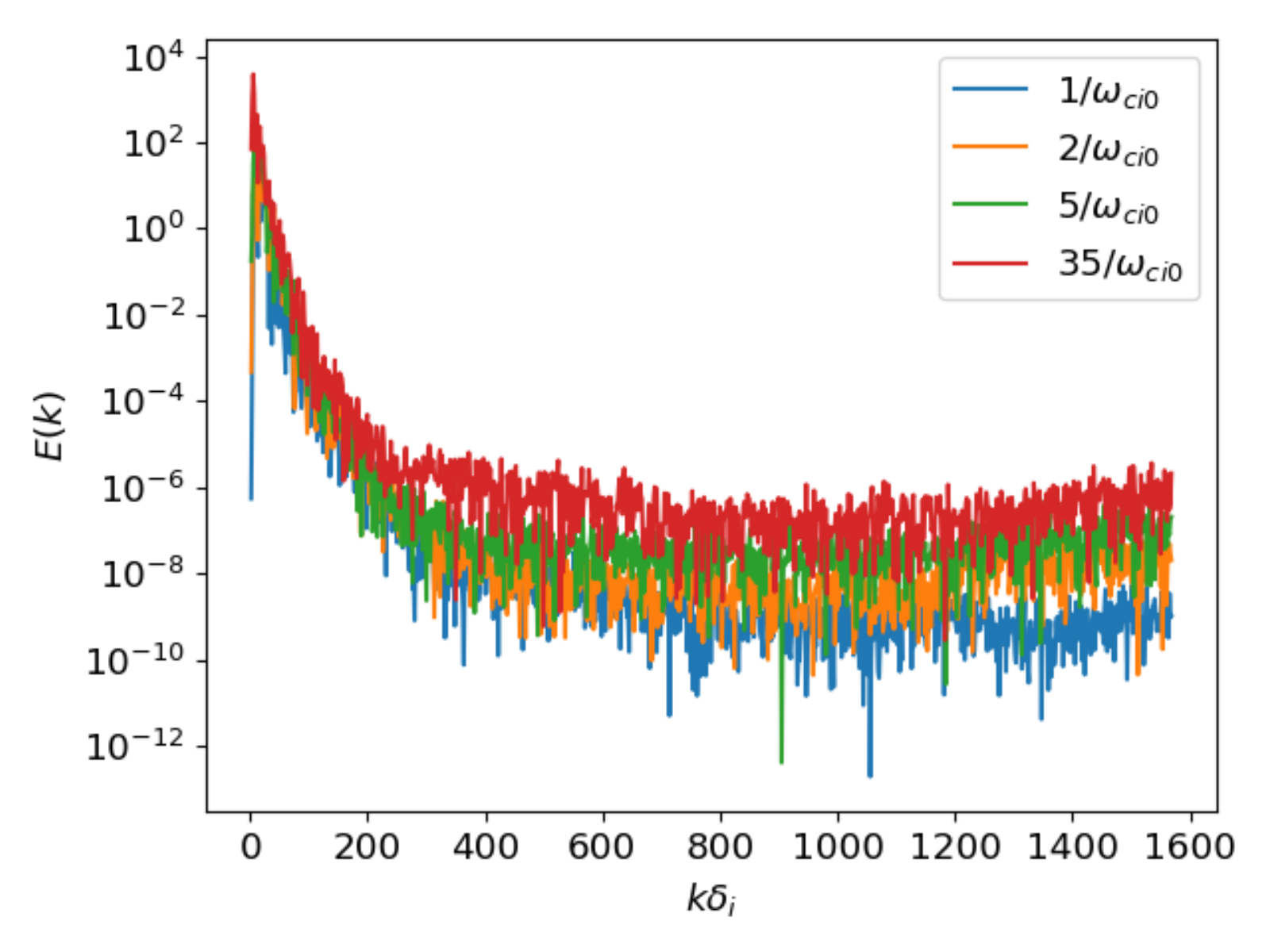}
  \caption{Power spectral density as a function of wavenumber for
    ion-to-electron mass ratios $M=25$ (left) and $M=1836$ (right)
    using the using the 10M model.  Solutions are shown for several
    different times during the evolution.  The ion density is
    integrated in the $x$-direction to obtain the spectrum in the
    $y$-direction.  Note that the energy in all modes late-in-time,
    including higher-wavenumber modes, is higher for simulations that
    use a realistic mass ratio compared to a mass ratio of $25$.}
  \label{fig:10Mfft}
\end{figure*}
For the 10M model, the electron inertia has a significant effect on
RTI evolution, particularly in the late, nonlinear phase of the
instability.  The RTI simulations presented in
Refs.[\onlinecite{huba_1996, huba_1998}] use hybrid models to
incorporate FLR effects with the ions evolved kinetically and a
massless electron fluid.  This recovers the correct asymptotic physics
in the limit of realistic ion-to-electron mass ratio.  However, recent
kinetic simulations of RTI \cite{umeda_2016, umeda_2017} use an
ion-to-electron mass ratio of $25$ which could produce inaccurate
nonlinear RTI evolution.  The kinetic effects in these results, for
example through the role of FLR effects in the RTI evolution, could
significantly over-predict the stabilized modes and the nonlinear
dynamics with an artificial ion-to-electron mass ratio.
Figure~\ref{fig:10Mdens} presents ion density using the 10M model for
the same multimode simulation presented in Sec.~\ref{sec:5Mmassratio}
for early- and late-time evolution using the same ion-to-electron mass
ratios of $25$ and $1836$.  

Note that there are significant differences between the two mass
ratios with the 10M model both early-in-time and late-in-time.  The
realistic mass ratio evolves to a greater bubble/spike propagation
(amplitude) compared to the solution that uses a mass ratio of $25$.
Additionally, the qualitative results of Fig.~\ref{fig:10Mdens}
clearly show that higher wavenumbers are dominant when using the
realistic mass ratio.  This is further noted from the power spectral
density as a function of wavenumber in the $y$-direction plotted for
several different times in Fig.~\ref{fig:10Mfft}.  The ion density is
used to create this plot and the density is integrated in the
$x$-direction to obtain a domain averaged quantity describing the
spectrum in the $y$-direction.  A closer look at the power spectral
density after $35/\omega_{ci0}$ shows that the realistic mass ratio
10M results have greater energies for high-wavenumber modes compared
to results using the 10M model with a mass ratio of $25$.  In fact,
even as early as $5/\omega_{ci0}$, the power spectral density differs
substantially for the two mass ratio results, particularly for
high-wavenumber modes. A mass ratio of $25$ over-predicts the
stabilization by damping out high-wavenumber modes late-in-time
compared to a mass ratio of $1836$.  This has implications for RT
turbulence studies, when FLR effects and kinetic effects become
important, as artificially higher electron mass may not capture
high-wavenumber spectra accurately.

\begin{figure*}[!htb]
  \includegraphics[width=0.45\linewidth]{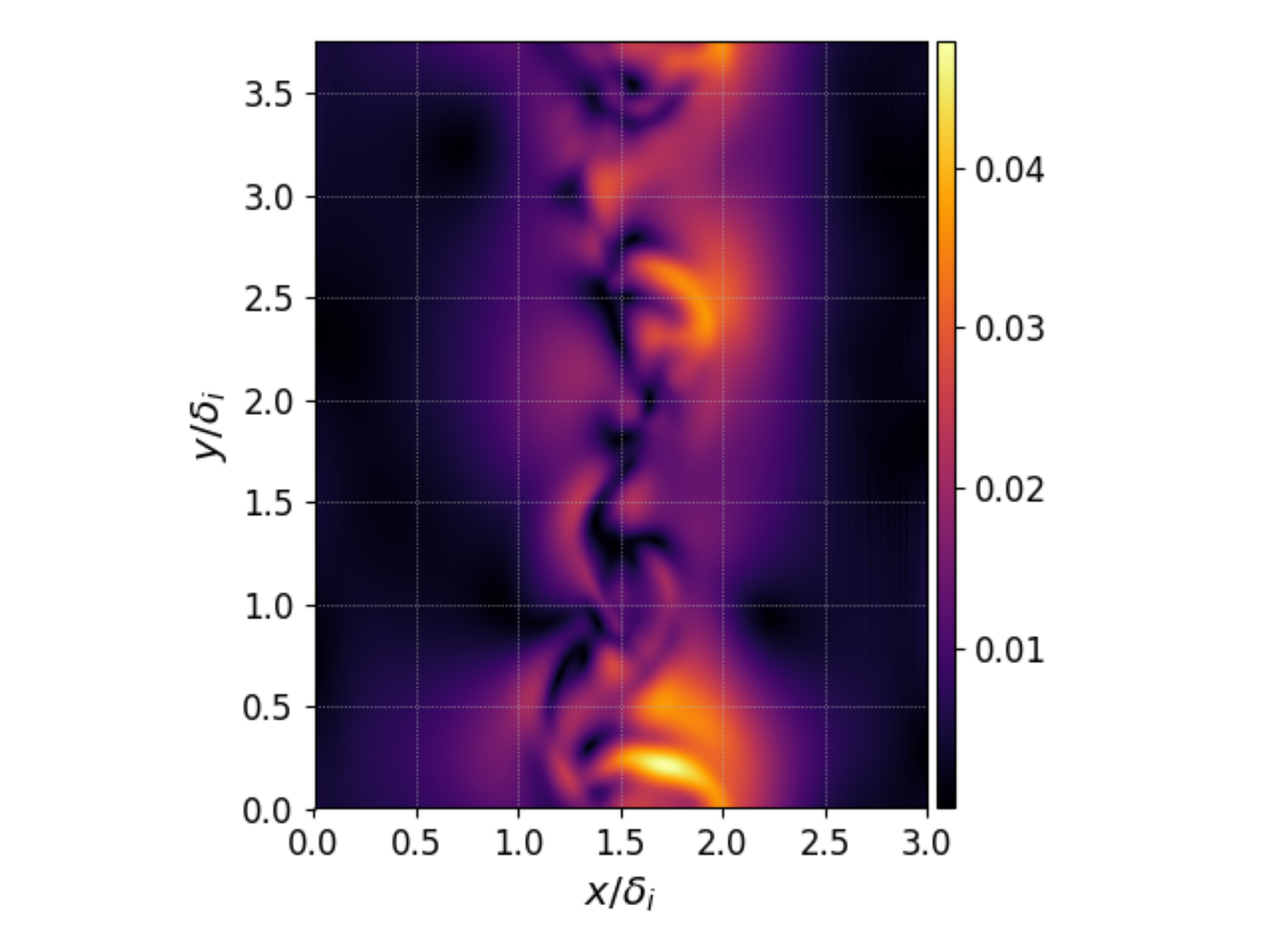}
  \includegraphics[width=0.45\linewidth]{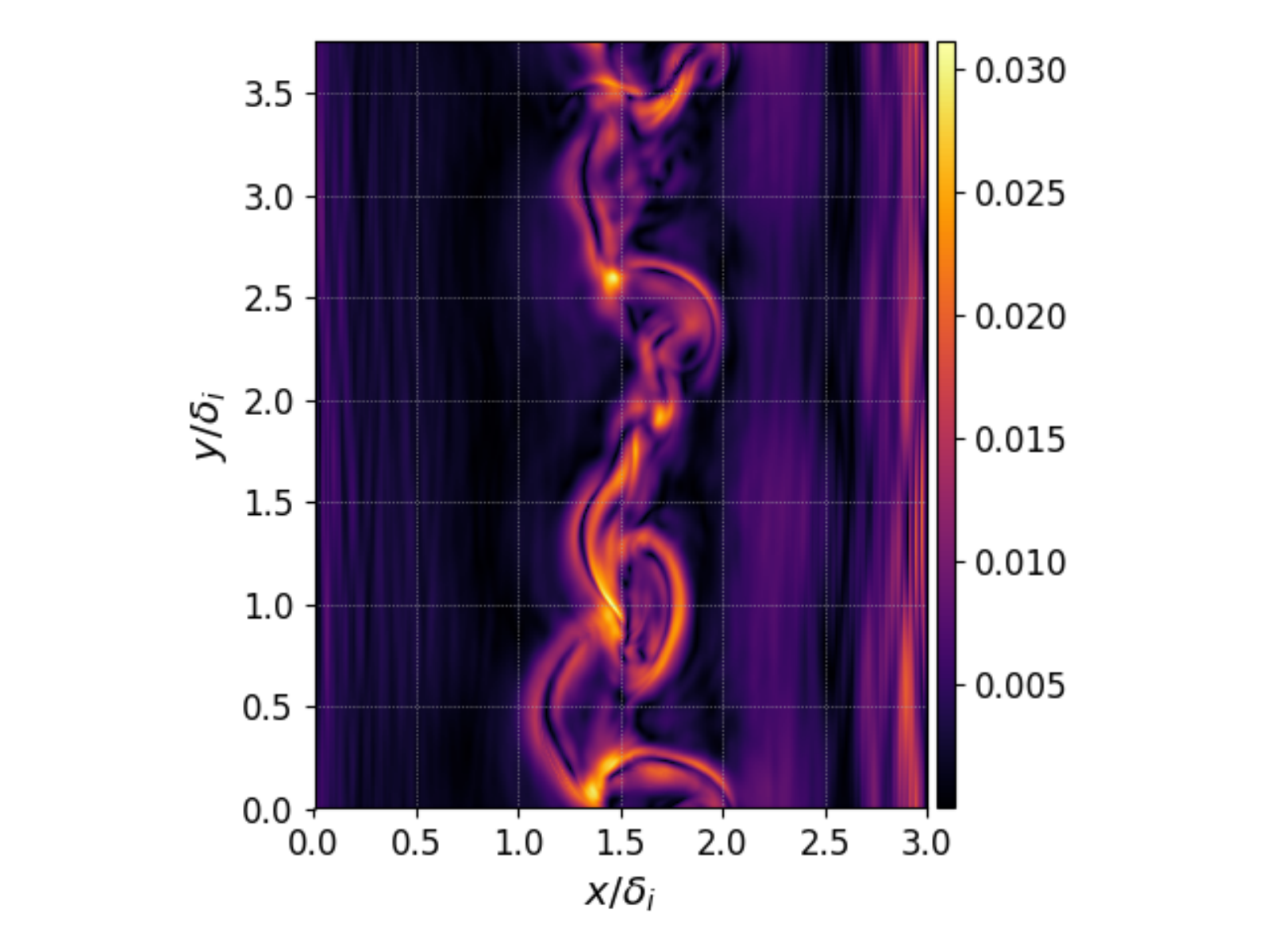}
  \includegraphics[width=0.45\linewidth]{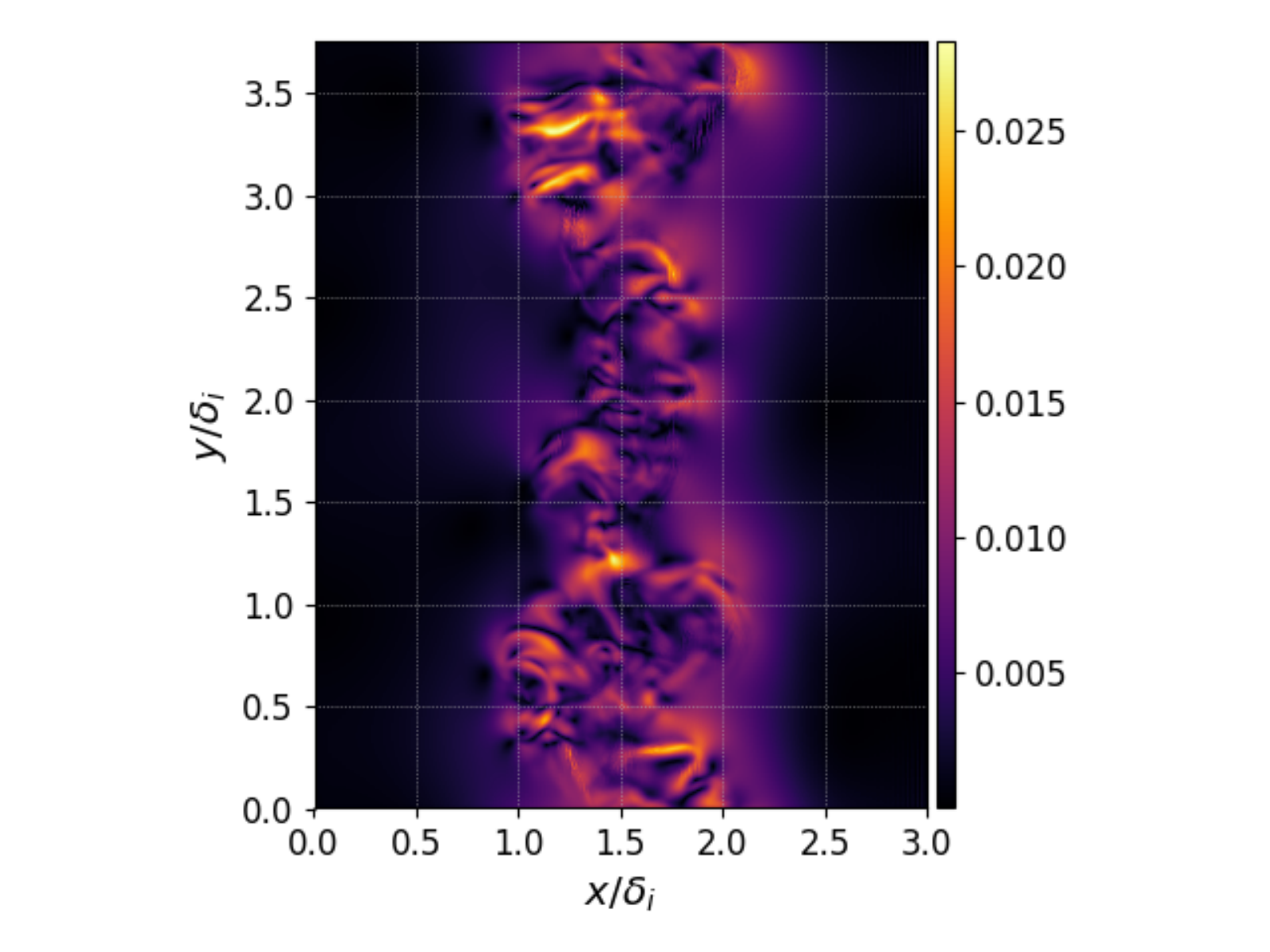}
  \includegraphics[width=0.45\linewidth]{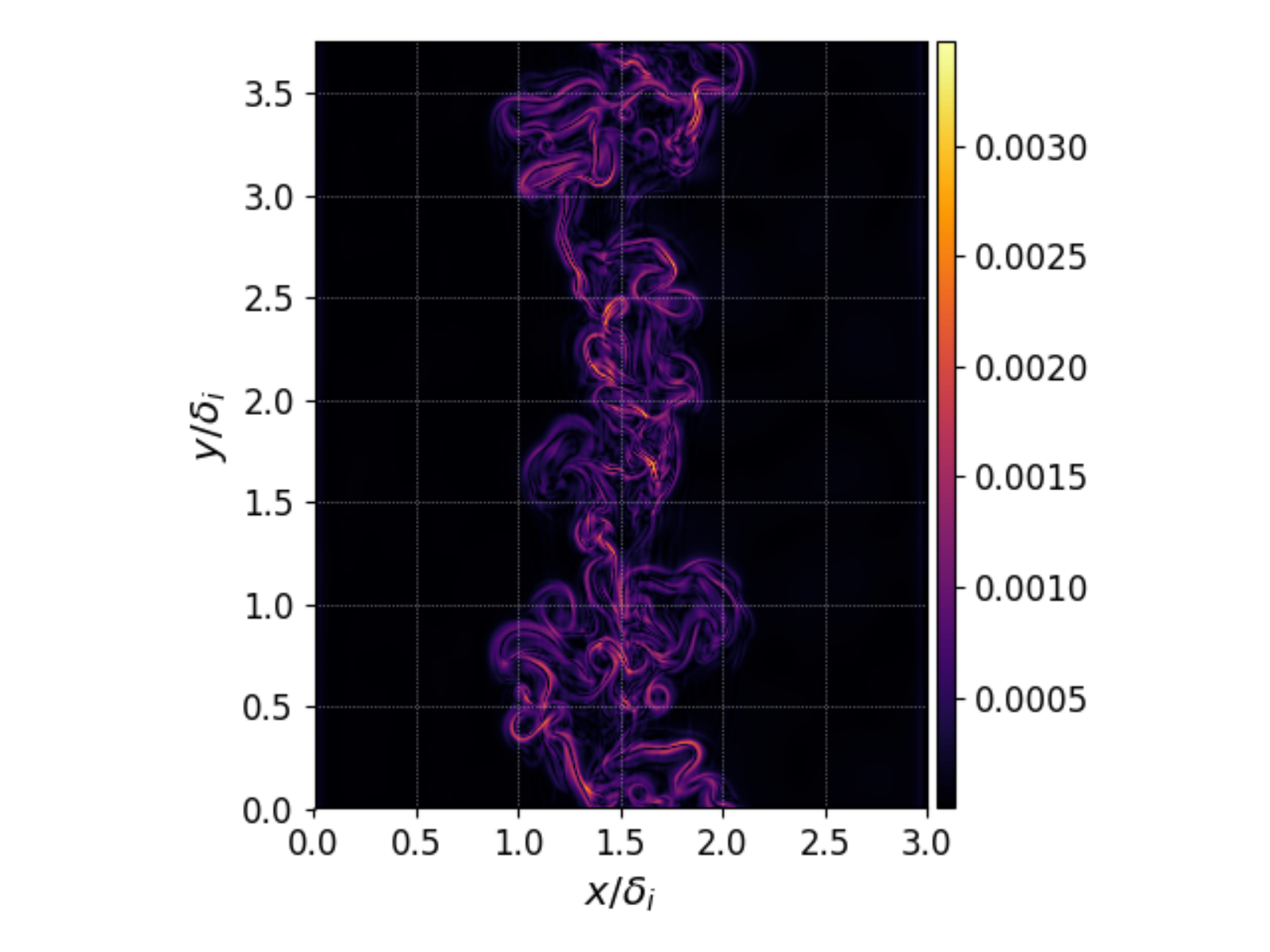}
  \caption{Ion (left) and electron (right) agyrotropy after
    35/$\omega_{ci0}$ for ion-to-electron mass ratios $M=25$ (top) and
    $M=1836$ (bottom) using the 10M model.  Note that the ion
    agyrotropy is approximately a factor of two lower and the electron
    agyrotropy is an order of magnitude lower when using a mass ratio
    of $1836$ compared to a mass ratio of $25$.}
  \label{fig:10MdiffMagyro}
\end{figure*}
To address the source of differences between the 10M solutions using
different mass ratios, it is illuminating to look at the ion and
electron agyrotropy for these results.  Figure~\ref{fig:10MdiffMagyro}
presents the agyrotropy described by Eq. \ref{eq:agyro} for the 10M
simulations presented in Fig.~\ref{fig:10Mdens} for $35/\omega_{ci0}$
using different ion-to-electron mass ratios.  Note that there is a
clear and significant difference in the agyrotropy when using
different mass ratios.  The top left plot, which presents ion
agyrotropy for a mass ratio of $25$ is approximately a factor of two
larger in terms of magnitude of agyrotropy compared to the bottom left
plot which presents ion agyrotropy for a mass ratio of $1836$.  This
implies that ion FLR effects are stronger when using a lower mass
ratio, and consequently the FLR stabilization is greater.  Similarly,
the top right plot, which presents electron agyrotropy for a mass
ratio of $25$ is larger, by an order of magnitude, than the bottom
right plot which presents electron agyrotropy for a mass ratio of
$1836$.  This indicates that the electron FLR effects are
over-predicted when using an artificially small ion-to-electron mass
ratio.  Despite the electron agyrotropy having larger differences for
the different mass ratios compared to the ion agyrotropy, ``hybrid''
simulations performed using 5M electrons and 10M ions reveal that the
electron FLR effects do not affect the nonlinear RT dynamics and it is
the ion FLR effects that dominate in the FLR stabilization noted in
these simulations.

\subsection{Comparison of 5M and 10M results}

\begin{figure}[!htb]
  \includegraphics[width=1.0\linewidth]{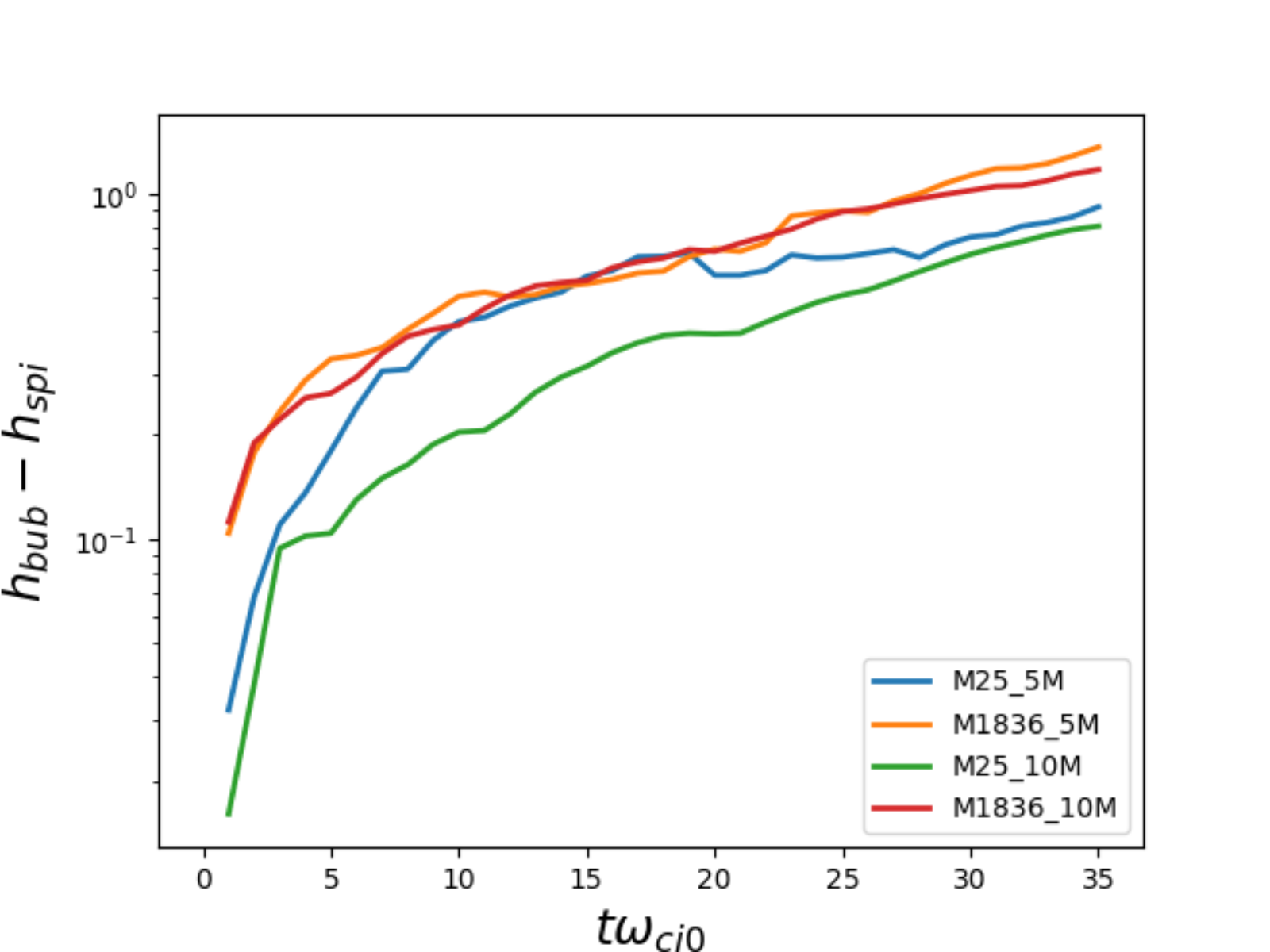}
  \caption{Bubble/spike propagation for each of the 5M and 10M models
    each using mass ratios of $25$ and $1836$.  }
  \label{fig:growth}
\end{figure}
The bubble/spike propagation is plotted in Fig.~\ref{fig:growth} for
the 5M and 10M models each using ion-to-electron mass ratios of $25$
and $1836$.  Note that, while the propagation for the different mass
ratios for the 5M model converges around $15-20/\omega_{ci0}$, it
begins to diverge again as the simulation evolves later in time.  This
may be due to an increased amount of secondary Kelvin-Helmholtz
instability in the lower mass ratio simulations compared to realistic
mass ratio simulations as observed from Fig.~\ref{fig:5Mdens}.  The
10M model however, has significant differences in the bubble/spike
propagation for all times.  In addition to the lower ion-to-electron
mass ratio damping higher-wavenumber modes late in time for the 10M
model, the lower mass ratio 10M results also have slower bubble/spike
propagation throughout compared to the realistic mass ratio results.

\begin{figure}[!htb]
  \includegraphics[width=1.0\linewidth]{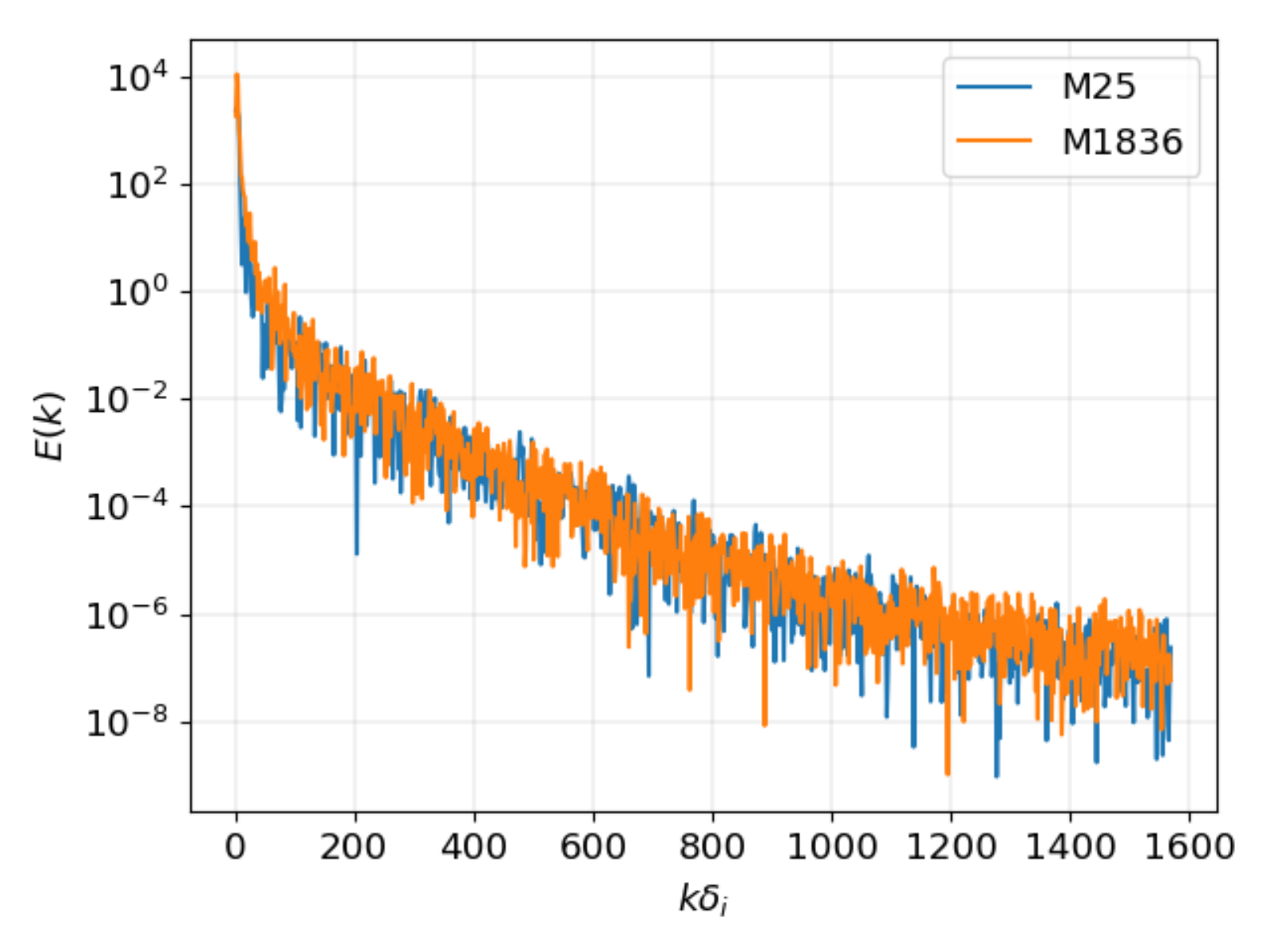}
  \caption{Power spectral density for the 5M model corresponding to a
    time of 35/$\omega_{cio}$ from Fig. \ref{fig:5Mfft} for both mass
    ratios of $25$ and $1836$.  Note that the two ion-to-electron mass
    ratio simulations produce very similar power spectra.}
  \label{fig:5Mfft1d}
\end{figure}
\begin{figure}[!htb]
  \includegraphics[width=1.0\linewidth]{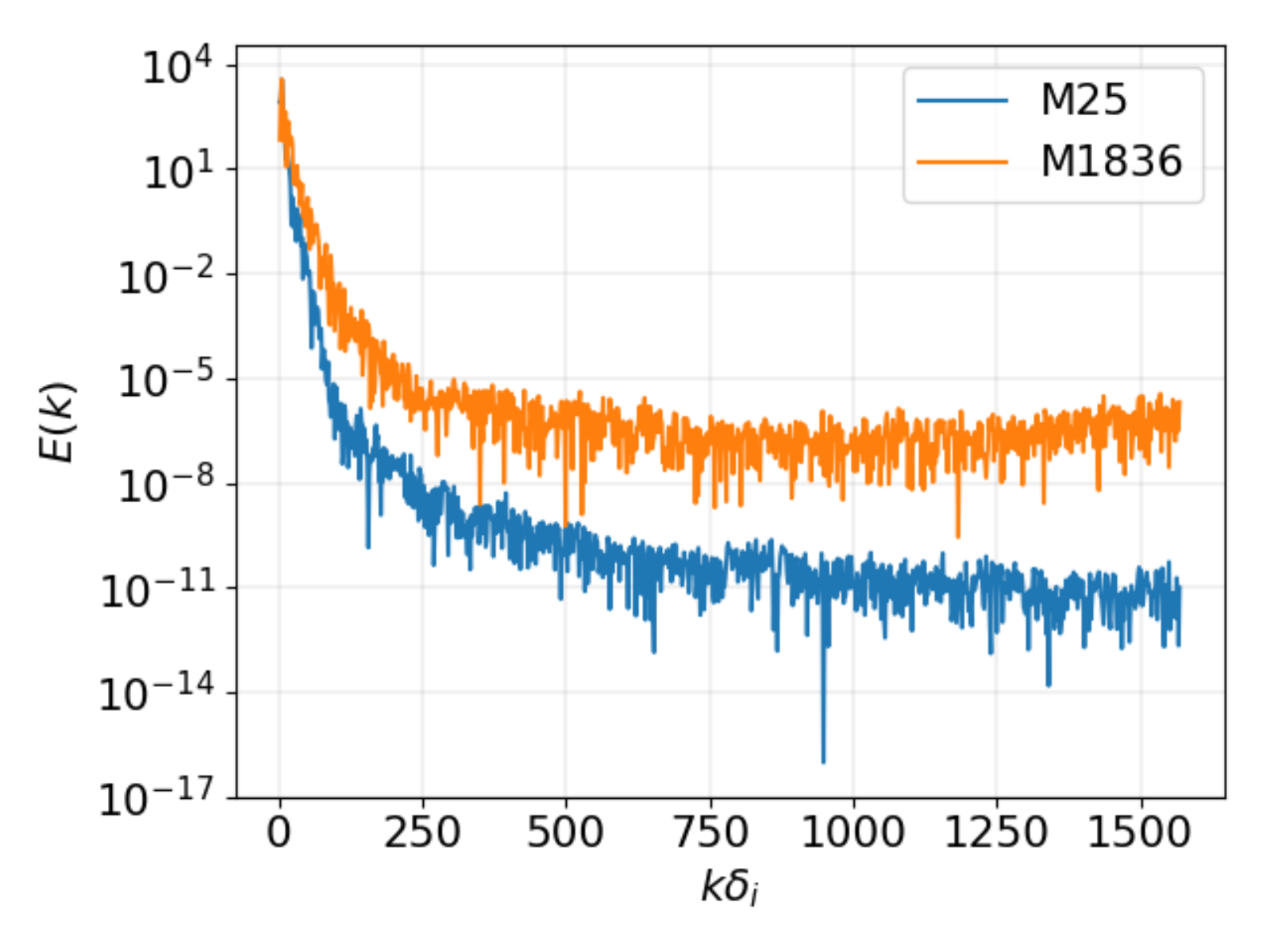}
  \caption{Power spectral density for the 10M model corresponding to a
    time of 35/$\omega_{cio}$ from Fig. \ref{fig:10Mfft} for both mass
    ratios of $25$ and $1836$. Note that the lower ion-to-electron
    mass ratio simulations have consistently lower power spectral
    density compared to the higher mass ratio.}
  \label{fig:10Mfft1d}
\end{figure}
Figure~\ref{fig:5Mfft1d} plots the 5M model power spectral density as
a function of wavenumber ($k_y$) at a time of $35/\omega_{ci0}$ (from
Fig.~\ref{fig:5Mfft}) for each of the two mass ratios. Similarly,
Fig.~\ref{fig:10Mfft1d} plots the 10M model power spectral density as
a function of wavenumber ($k_y$) at the same time (from
Fig.~\ref{fig:10Mfft}) for each of the two mass ratios.  To reiterate,
the 5M model (Fig.~\ref{fig:5Mfft1d}) produces very similar power
spectral densities for late-time RT for the different mass ratios.
However, for the 10M model (Fig.~\ref{fig:10Mfft1d}), the artificially
high electron mass (lower ion-to-electron mass ratio) underpredicts
the power spectral density by a substantial amount (3-5 orders of
magnitude) for all wavenumbers present in the spectrum and therefore
over-predicts the FLR stabilization.  Lastly, the significant
differences in RTI evolution between the 5M and 10M results for both
mass ratios are due to the absence of FLR effects in the 5M model.

\section{Summary}

This work studies the role of ion/electron FLR effects and the role of
electron inertia in the nonlinear phase of the magneto-RTI.  A
five-moment two-fluid model, that captures Hall effects but does not
include FLR effects, is compared to a ten-moment two-fluid model which
contains Hall and FLR effects for different ion-to-electron mass
ratios.  The simulations are performed in a regime where FLR effects
are considered to be important.

The results presented here show that early-time RTI growth has slower
bubble/spike propagation for a lower ion-to-electron mass ratio
compared to a realistic mass ratio for both the five- and ten-moment
models.  Late-in-time, well into the nonlinear phase of the
instability, the five-moment results with different mass ratios
converge to similar bubble/spike propagation and similar dominant
wavenumber spectra.  However, the late-time nonlinear dynamics of the
ten-moment model exhibit significant differences for the different
ion-to-electron mass ratios.  For a lower ion-to-electron mass ratio
(i.e. with artificially large electron inertia) the bubble/spike
propagation is slower than for a realistic mass ratio.  Furthermore,
the lower mass ratio ten-moment results damp the higher-wavenumber
modes that are present in the realistic mass ratio results.  Exploring
the ion and electron agyrotropy shows that the lower mass ratio
over-predicts both ion and electron FLR effects compared to the
realistic mass ratio results which is responsible for the differences
observed in the ten-moment model.  However, it is the ion FLR effects
that affect the FLR stabilization of high-wavenumber RT in these
results for the different mass ratios.

Previous work\cite{ng2017} has compared the agyrotropy between a
ten-moment model and a fully kinetic model to find that ten-moment
models produce lower values of agyrotropy compared to kinetic results
but the values are close in magnitude.  This implies that the
ten-moment results presented here are valid for understanding
implications when FLR effects are important.  The ten-moment model is
used here for reasons of computational efficiency as compared to a
well-resolved 2X3V (2 spatial dimensions and 3 velocity dimensions)
fully kinetic model that would otherwise be necessary for such a
study.  The results presented here highlight the role of electron
inertia in RTI evolution, particularly in the nonlinear phase of the
instability, and have implications for problems where FLR and other
kinetic effects may be important.

\begin{acknowledgments}
  The authors acknowledge Advanced Research Computing at Virginia Tech
  for providing computational resources and technical support that
  have contributed to the results reported within this paper
  (\url{http://www.arc.vt.edu}).  Petr Cagas is acknowledged for his
  assistance with post-processing by developing and supporting
  Postgkyl (\url{http://gkyl.readthedocs.io}), a valuable analysis and
  plotting tool.  The work presented here was supported by the
  Department of Energy High-Energy-Density Laboratory Plasmas program
  under grant number DE-SC0016515.
\end{acknowledgments}

\section{\label{sec:reference}Reference}

\end{document}